\documentclass[sigconf]{acmart}
\def\BibTeX{{\rm B\kern-.05em{\sc i\kern-.025em b}\kern-.08emT\kern-.1667em\lower.7ex\hbox{E}\kern-.125emX}}
    
\copyrightyear{2020}
\acmYear{2020}
\setcopyright{acmcopyright}
\acmConference[ACMSE 2020]{2020 ACM Southeast Conference}{April 2--4, 2020}{Tampa, FL, USA}
\acmBooktitle{2020 ACM Southeast Conference (ACMSE 2020), April 2--4, 2020, Tampa, FL, USA}
\acmPrice{15.00}
\acmDOI{10.1145/3374135.3385272}
\acmISBN{978-1-4503-7105-6/20/03}

\def\BibTeX{{\rm B\kern-.05em{\sc i\kern-.025em b}\kern-.08emT\kern-.1667em\lower.7ex\hbox{E}\kern-.125emX}}
    
\usepackage{flushend} 
\usepackage{geometry}
\hypersetup{draft}
\usepackage{balance}
\begin{document}

%
% The "title" command has an optional parameter, allowing the author to define a "short title" to be used in page headers.
% \title{LSTM Based Deep Learning Approach for Repayment Prediction of Peer-to-Peer (P2P) Lending Market}
% \title{Incorporation of the Macroeconomic Factor into LSTM Approach for Repayment Prediction of Peer-to-Peer (P2P) Lending Market}
\title{Improving Investment Suggestions for Peer-to-Peer (P2P) Lending via Integrating Credit Scoring into Profit Scoring}

%
% The "author" command and its associated commands are used to define the authors and their affiliations.
% Of note is the shared affiliation of the first two authors, and the "authornote" and "authornotemark" commands
% used to denote shared contribution to the research.
\author{Yan Wang}

\orcid{1234-5678-9012}
\affiliation{%
  \institution{Kennesaw State University}
  \city{Kennesaw, GA}
  \state{USA}
}
\email{ywang63@students.kennesaw.edu}

\author{Xuelei Sherry Ni}
\affiliation{%
  \institution{Kennesaw State University}
  \city{Kennesaw, GA}
  \state{USA}
}
\email{sni@kennesaw.edu}

% \affiliation{%

% \affiliation{%
 
% \affiliation{%

% \affiliation{%



%
% By default, the full list of authors will be used in the page headers. Often, this list is too long, and will overlap
% other information printed in the page headers. This command allows the author to define a more concise list
% of authors' names for this purpose.
\renewcommand{\shortauthors}{Wang and Ni.}

%
% The abstract is a short summary of the work to be presented in the article.
\begin{abstract}
In the peer-to-peer (P2P) lending market, lenders lend the money to the borrowers through a virtual platform and earn the possible profit generated by the interest rate.
From the perspective of lenders, they want to maximize the profit while minimizing the risk. 
Therefore, many studies have used machine learning algorithms to help the lenders identify the ``best" loans for making investments. 
The studies have mainly focused on two categories to guide the lenders' investments: one aims at minimizing the risk of investment (i.e., the credit scoring perspective) while the other aims at maximizing the profit (i.e., the profit scoring perspective). 
However, they have all focused on one category only and there is seldom research trying to integrate the two categories together.
Motivated by this, we propose a two-stage framework that incorporates the credit information into a profit scoring modeling.
We conducted the empirical experiment on a real-world P2P lending data from the US P2P market and used the Light Gradient Boosting Machine (lightGBM) algorithm in the two-stage framework. 
Results show that the proposed two-stage method could identify more profitable loans and thereby provide better investment guidance to the investors compared to the existing one-stage profit scoring alone approach.
Therefore, the proposed framework serves as an innovative perspective for making investment decisions in P2P lending. 
\end{abstract}

%
% Keywords. The author(s) should pick words that accurately describe the work being
% presented. Separate the keywords with commas.

%
% A "teaser" image appears between the author and affiliation information and the body 
% of the document, and typically spans the page. 
%   \Description{Enjoying the baseball game from the third-base seats. Ichiro Suzuki preparing to bat.}

%
% This command processes the author and affiliation and title information and builds
% the first part of the formatted document.
\maketitle

\section{Introduction} \label{introduction}
Peer-to-peer (P2P) lending consists of the practice of matching anonymous lenders with borrowers through an electronic platform so lenders could directly invest on (lend to) certain borrowers \cite{bachmann2011online}. 
In general, lenders could earn higher returns relative to savings and other investment products offered by banking when borrowers pay back their loans as scheduled.
However, the loans on the P2P market are unsecured and investors need to tolerate the risk of losing part or even all of their principal if borrowers default the loans. 
To help investors find out the safer loans with the relatively lower risk, it is beneficial to evaluate each loan from the perspective of ``the risk level", which is typically done by estimating the probability of default (PD). 
Loans with lower PDs are considered safer than those with higher PDs and vice versa. 
The PD for each loan can be predicted by considering its characteristics, such as the loan amount, the loan purpose, the assets of the borrowers, etc.
The above-mentioned approach is known as the credit scoring approach, which poses a classification problem that classifies the loans into either (1) the default case if the predicted PD exceeds a certain predefined threshold, or (2) the non-default case otherwise.
Subsequently, the credit scoring approach recommends lenders to invest in non-default loans or the loans with lower predicted PDs because of the potentially lower risk.

In the P2P market, minimizing the risk is one but not the only objective for investors. 
The profit gain of the loan even matters more for lenders, making it crucial to evaluate each loan at ``the profit level", which is known as the profit scoring approach. 
In \cite{serrano2016use}, profit scoring was first proposed as an alternative to credit scoring in P2P lending and the internal rate of return (IRR) was used as the measure of the profit. 
IRR is a well-known financial formula \cite{brealey2012principles}.
For example, suppose there are two borrowers obtaining a \$100 loan each, and suppose the first borrower pays back \$150 and the second one pays back \$95. 
Then the IRRs for the first and the second borrowers are $\frac{\$150-\$100}{\$100}=50\%$ and $\frac{\$95-\$100}{\$100}=-5\%$, respectively. 
% In \cite{serrano2016use}, 
The profit scoring approach poses a regression problem that predicts the IRR for each loan and the loans with a high predicted IRR are the good candidates for investors.
Later on, the authors in \cite{xia2017cost} pointed out that the annualized rate of return (ARR), rather than IRR, is a more appropriate measure of profit.
% Especially in P2P lending, the duration of the repayments varies for different loans, making it unsuitable to use IRR to make comparisons. 
This is due to the various duration of the repayments for different loans.
Instead, ARR takes the true term of loans into account thus is more appropriate to evaluate the loans with different repayment duration.
The ARR formula is described in Equation \ref{ARR_equation}, where Pa is the total payment when the loan expires, Pr denotes the principal, and Y is the number of years of the repayment process. 
Again, suppose there are two borrowers obtaining a \$100 loan each, and suppose the first borrower pays back \$150 in 1 year while the second borrower pays \$150 in 2 years. 
Both of them generate the IRR valued $50\%$. 
However, the ARRs for the first and the second borrowers are $(\frac{\$150}{\$100})^{(1/1)}= 1.5$ and $(\frac{\$150}{\$100})^{(1/2)}= 1.2$, respectively. 
The investment gains the profit in a shorter period of time is valued higher by ARR. 
Considering that in P2P lending, the duration of the repayments varies for different borrowers, we will use ARR as the profit measurement in our study.
%Subsequently, the profit scoring approach recommends lenders to invest in the loans with higher predicted ARRs because potentially they will generate higher returns. 

\begin{equation}\label{ARR_equation}
ARR = (\frac{Pa}{Pr})^{1/Y}
\end{equation}

Both of the credit scoring approach and the profit scoring approach can be used to evaluate loans and make recommendations to investors.
However, they work from different perspectives.
As pointed out in \cite{serrano2016use}, the factors determining the profit differ from those determining the PD, although overlapping factors exist. 
The credit scoring approach helps lenders minimize the potential default risk.
It identifies the loans with lower PDs and these ``safe" loans are considered as the ``good" loans. 
From the credit scoring perspective, the ``safe'' loans may lead to a good profit since they have a higher probability of being fully repaid. 
On the other hand, the profit scoring method identifies the loans with higher predicted profits based on the condition that borrowers fully pay off their loans (e.g., they are non-default loans) and these ``more profitable" loans are considered as ``good" loans from the profit scoring perspective.
Although working from different perspectives, the final objective of both credit scoring and profit scoring is to help investors get more profit from the investment.

Considering that credit scoring only focuses on PDs and totally ignores the profit while profit scoring only targets on the profit and totally ignores the default risk, none of the two approaches could evaluate the loans comprehensively. 
It is intuitive that the higher PD the loan has, the higher interest rate it associates with, thus the higher profit it may lead to. 
Therefore, the credit scoring information may provide some additional power to the prediction of the profit and integrating the two scoring approaches may provide a better investment suggestion. 
Motivated by the aforementioned conjecture, we design a two-stage framework that could integrate the credit scoring information into the profit scoring method in the evaluation of loans.
To be specific, in stage 1, each loan's PD is estimated by a classifier.  The predicted PD then serves as an additional predictor in stage 2, where a regressor is used to get the predicted profit of each loan. 
Subsequently, the lenders might be able to select the loans with a higher predicted profit than those selected through the single-step approach. 

To our best knowledge, the proposed two-stage framework is the first study aiming at incorporating credit scoring and profit scoring together to evaluate loans. 
To validate the effectiveness of the proposed approach, we conducted an empirical study using a real-world data from Lending Club, which represents most of the P2P transactions in US. 
The results indicate that the two-stage approach outperforms the existing one-stage profit scoring alone approach with respect to the identification of the more profitable loans. 

This paper has been structured as follows. 
We will first review the related work of credit scoring and profit scoring in the P2P domain in Section \ref{relatedwork}. 
Section \ref{algorithms} gives a brief overview of the proposed two-stage modeling approach based on the Light Gradient Boosting Machine (lightGBM) algorithm. 
The details of the empirical study are further presented in Section \ref{experimentdesign}.
Section \ref{resultsanddiscussions} displays the 
experimental results. 
Conclusions and discussion are finally addressed in Section \ref{conclusion}.

\section{Related Work} \label{relatedwork}
In this section, the related work in credit scoring and profit scoring are summarized. 

\subsection{Research on Credit Scoring} \label{creditscoring}
In the P2P market, credit scoring is formulated as a classification problem with a binary outcome: default loans (i.e., more than 150 days past due) and non-default loans (i.e., fully paid). 
Different classifiers have been used in the credit scoring area, including logistic regression, support vector machine, Naive Bayes, k-nearest neighbors, random forest, and neural network \cite{polena2018determinants}.
Logistic regression is considered a natural method for credit scoring because of its relatively strong performance. 
Furthermore, it was shown that logistic regression could reach the best precision compared to other classifiers including support vector machine, Naive Bayes, and random forest on the Lending Club data \cite{kumar2016credit}.
In \cite{malekipirbazari2015risk}, a random forest based classification approach was used to identify the loan status and it turned out the random forest model could reach a higher accuracy than support vector machine or logistic regression.
In \cite{kim2019predicting}, a deep dense convolutional network was created to predict the repayment amount of P2P lending.
Tree-based ensemble algorithms including lightGBM and XGBoost methods have been used to evaluate the loans on the Lending Club platform as well \cite{ma2018study}.
Moreover, there have been some studies focusing on creating a hybrid model that aims to further improve the performance of the credit scoring approach.
For instance, in \cite{fu2017combination}, a hybrid model combining random forest and neural networks was proposed. 
Regardless of the various machine learning models proposed in the credit scoring area, all of them focused on targeting the ``safest" loans and totally ignore their profitability. 

\subsection{Research on Profit Scoring}
Recently, many studies have changed their focus from credit scoring to profit scoring. 
However, there is still limited research focusing on profit scoring for P2P lending. 
As discussed in Section \ref{introduction}, IRR and ARR have been used as the target for this approach \cite{serrano2016use}\cite{xia2017cost}.
Since both IRR and ARR are continuous, profit scoring is formulated as a regression problem. 
In \cite{serrano2016use}, multiple linear regression and decision tree models are used for the prediction of IRR. 
% Later on, authors in \cite{xia2017cost} pointed out that IRR is not an appropriate profitability measure because of the inherent characteristics of P2P lending. 
% For P2P loans, the true term of loans varies because of early repayments, which will lead to less return compared to repaying gradually on time. 
% However, IRR does not consider the repayment duration in reality.
% Instead, ARR, which considers the true repayment term of the loan, is more optimal than IRR when being used for the profitability measure. 
% In \cite{serrano2016use}, the measure of the profitability of loans is first introduced to the P2P area and authors use IRR as a measure of profitability.
% Since IRR is continuous, profit scoring is formulated as a regression problem. 
% Multiple linear regression and decision tree models are used for the prediction of IRR. 
% Later on, authors in \cite{xia2017cost} pointed out that IRR is not an appropriate profitability measure because of the inherent characteristics of P2P lending. 
% For P2P loans, the true term of loans varies because of early repayments, which will lead to less return compared to repaying gradually on time. 
% However, IRR does not consider the repayment duration in reality.
% Instead, ARR, which considers the true repayment term of the loan, is more optimal than IRR when being used for the profitability measure. 
In \cite{xia2017cost}, a cost-sensitive extreme gradient boosting (CSXGBoost) model is used to get the predictions of ARR. 
Regardless of the choice of the profit measure, profit scoring models only focus on finding the most ``profitable" loans and totally ignore their default risk. 

\section{The Proposed Two-stage Approach} \label{algorithms}
\begin{figure}[htbp]
\centering
\includegraphics[width=3.3in]{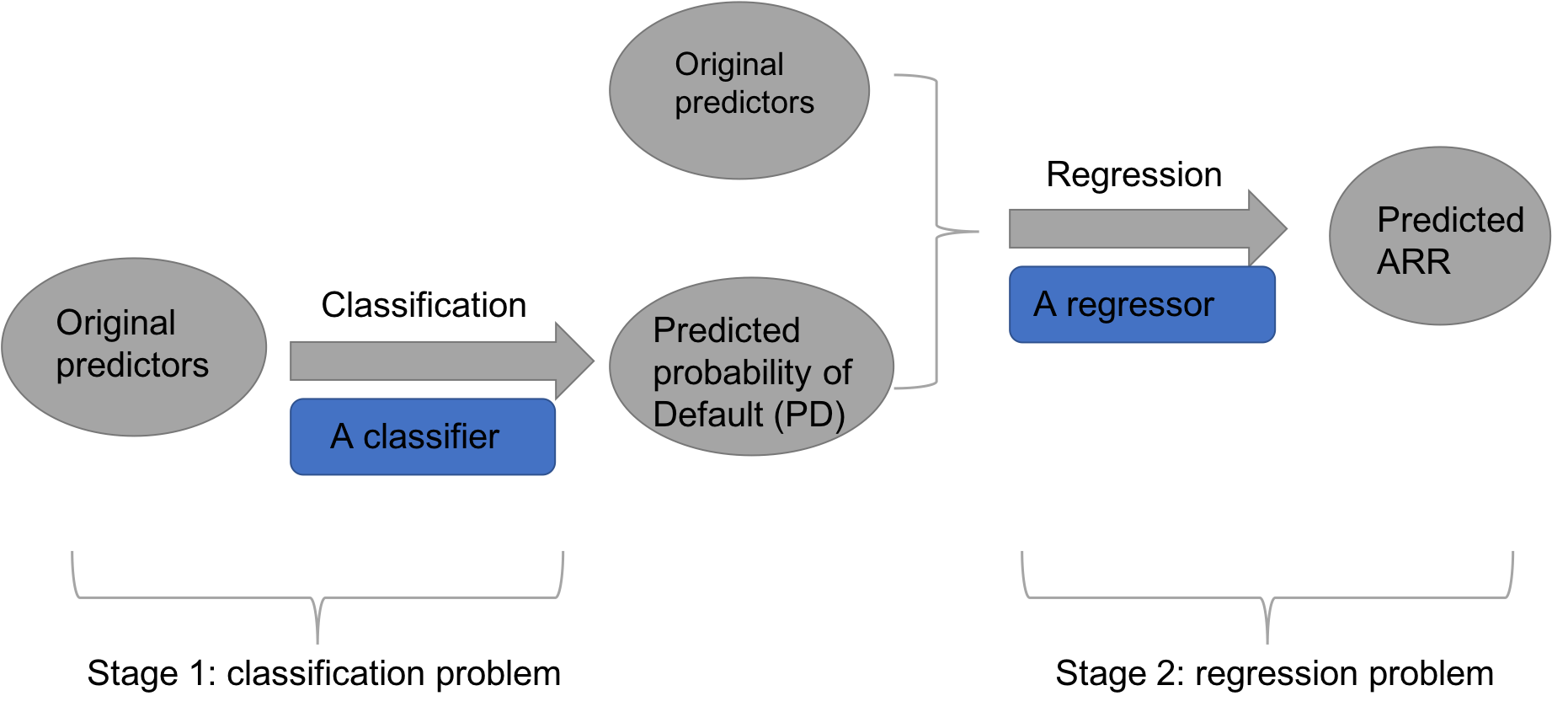}
\caption{The Illustrative Structure of the Two Stage Model}
\label{twostagemodel}
\end{figure}

% \subsection{Methodology} 
As discussed in Section \ref{introduction}, the credit scoring information may  be beneficial in the detection of more profitable loans. 
% As mentioned in Section \ref{relatedwork}, loan\_status is the target variable for the credit scoring approach while ARR is the target variable for the profit scoring approach.
% Furthermore, 
In order to incorporate the credit information into profit scoring, an intuitive approach is to use the loan status (i.e., default or non-default) as an additional predictor in the profit scoring approach. 
Although it works on the historical data, it cannot be used in real applications due to the lack of the value for the loan status when a loan is initiated and it is when lenders would like to assess its profitability.
To overcome the above-mentioned problem, a two-stage method is developed and its structure is shown in Figure \ref{twostagemodel}.
Stage 1 predicts PD by formulating it into a binary classification problem. 
The predicted PD generated from stage 1 is then used as an additional feature in stage 2 for the prediction of ARR.
The design of the two-stage approach is based on the assumption that the information of PD is predictive for ARR. 
We hope that adding PD as the additional predictor may help avoid the loans with extremely high profit while extremely high risk, which is especially helpful for conservative investors. 

As shown in Figure \ref{twostagemodel}, one classifier and one regressor are needed in stage 1 and stage 2, respectively. 
Theoretically, all kinds of classifiers and regressors could be used in the two-stage modeling process.
% Since the purpose of our experiment is to confirm that incorporating the credit information into profit scoring could help the identification of the more ``profitable" loans compared to single profit scoring approach, the classifier in stage 1 and the regressor in stage 2 should be any models. 
In this study, we select lightGBM as both the classifier in stage 1 and the regressor in stage 2.

LightGBM originated from Gradient Boosting Decision Tree (GBDT), which is an ensemble learning approach using the decision tree as the base classifier. 
GBDT could enhance a weak classifier into a strong one by iterative training \cite{zhang2015gradient}.
It soon became a deadly weapon in many machine learning tasks and more than half of the championship programs in the Kaggle competitions used GBDT \cite{ma2018study}.
XGBoost is one type of GBDT proposed in 2015.
In recent years, XGBoost has been frequently applied because of its rapidness and scalability \cite{chen2015xgboost}. 
LightGBM, designed in 2016, is an additional novel type of GBDT and was proposed to solve the problems encountered by XGBoost in large-scale data.
Details of the lightGBM theory can be found in \cite{ke2017lightgbm}.
LightGBM supports efficient parallel training so it could have a lower computational cost while having better performance than XGBoost \cite{ke2017lightgbm}. 
As a result, LightGBM is becoming more preferred in sorting, classification, and regression tasks \cite{song2019prediction}.
As mentioned in Section \ref{creditscoring}, lightGBM was first introduced into the P2P area for the prediction of loan repayments \cite{ma2018study}.
However, there has been no research that uses lightGBM for the prediction of a loan's profitability. This is the first application of LightGBM in such area.

In summary, we chose lightGBM as both the classifier in stage 1 and the regressor in stage 2 for the reasons as follows:

\begin{itemize}
\item LightGBM can handle both classification and regression problems \cite{ma2018study}. 
Using the same model in stages 1 and 2 can simplify the model structure. 
\item GBDT is an ensemble method and the performance is significantly better than most of the conventional machine learning methods, which has been well demonstrated in previous studies \cite{friedman2001greedy}\cite{xie2009combination}\cite{jahrer2010combining}.
As one type of GBDT, lightGBM has shown to have good stability and accuracy \cite{zhang2019research}\cite{xiaojun2018empirical}. 
It has a relatively small computational cost but provides good training effect. 
\item This is the first attempt of using lightGBM in predicting the profitability of a loan.
\end{itemize}

Therefore, in our proposed two-stage lightGBM model, stage 1 is designed as a credit scoring model, which uses lightGBM to get the predicted PD for all the loans. 
The predicted PD is then used as an additional predictor in stage 2 for the prediction of ARR, which also uses the lightGBM algorithm. 
The hyper-parameters of the lightGBM model in both stages, including the number of trees, the number of levels for each tree, and the percentage of subsample used during each iteration, are tuned based on a trial and error approach with the goal of minimizing the loss on the testing set. 

\section{Empirical Study} \label{experimentdesign}
As discussed in Section \ref{introduction}, it is our hope that adding the credit scoring information would be beneficial in the detection of more profitable loans. 
Thus, in this study, we aim to answer the following research question explicitly based on P2P lending:

\textit{Is incorporating the credit information into the profit scoring approach better than the profit scoring alone approach in identifying the ``more profitable'' loans?}

To address the above-mentioned question, we design a comprehensive empirical study and the details along with the data used are described in the following subsections. 

\subsection{Data Source}
\begin{figure}[htbp]
\centering
\includegraphics[width=3.3in]{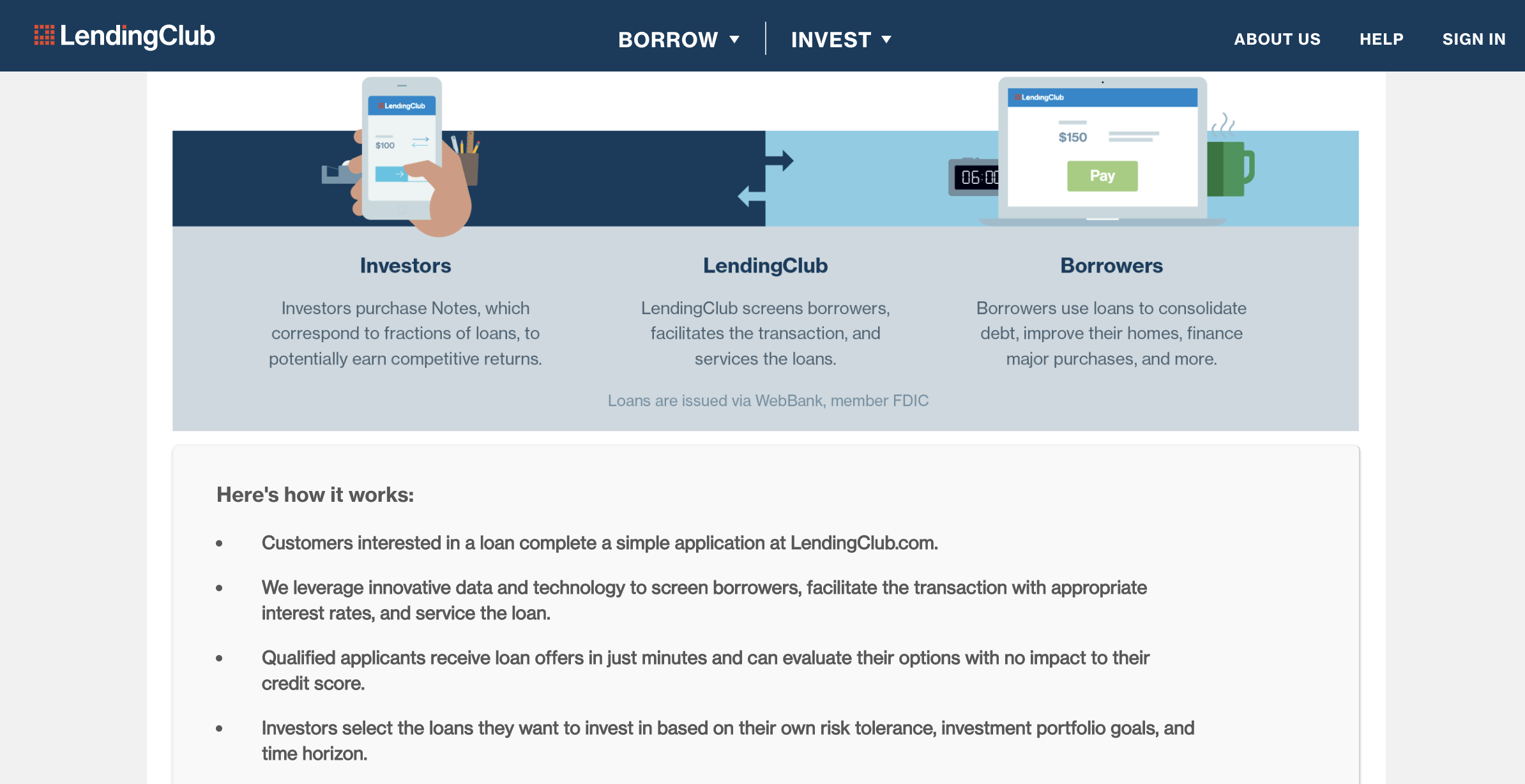}
\caption{The Homepage of the Lending Club Platform}
\label{lendingclubwebsite}
\end{figure}

The P2P lending market appeared in the US in February 2006.
By June 2012, Lending Club has become the largest P2P platform in the US with respect to the issued volume and the revenue.
Therefore, the transactions happened on the Lending Club platform are a good representative of the P2P market in the US.
Figure \ref{lendingclubwebsite} shows the homepage of the Lending Club website. 
The Lending Club acts as a third-party platform between the investors and the borrowers and the P2P transaction occurs when: (1) a borrower applies for a loan and Lending Club approves his/her applications;
and (2) an investor decides to invest on the loan if he/she thinks the borrower meets a certain criteria.

The historical Lending Club data can be acquired from its official website: \textit{https://www.lendingclub.com/info/download-data.action}. 
These data sets contain the information of millions of loan issued since 2007. The data is consistently updated with time going on and a newly updated data set is made available every quarter. 
Since most loans ($>$70\%) of Lending Club is 36-month long, we only study the loans that were issued before August 2016 (i.e., about 36 months before our analysis started). 
In this case, most loans are expired and have a clear state (e.g., not under the process of repayment). 
Furthermore, the loans that are still in the intermediate states (e.g., under the process of repayment) are filtered out. 
As a final result, we get the Lending Club data that leads to a total of 1,123,895 loans in our empirical study.

\subsection{Variables}
Each loan in the data is identified by the unique ID and the information of the loan is described by several features. 
Similar with previous research, these features are grouped into three categories: (1) loan characteristics; (2) credit worthiness; and (3) borrower information \cite{xia2017cost}.
After removing the features with high percentage of missing ($>70\%$ missing), we have 29 variables in the Lending Club data.
27 of the 29 variables are used as independent variables in the modeling stage. 
% They mainly fall into 3 categories: loan characteristics, credit worthiness, and borrower information. 
The definition of the variables belonging to each category can be found in the appendix. 
The remaining two variables are the targets: one for the credit scoring approach and the other for the profit scoring approach. 
We will introduce how they are generated in Section \ref{credittarget} and \ref{profittarget}, respectively.

\begin{figure}[htbp]
    \centering
    \includegraphics[width=3.3in]{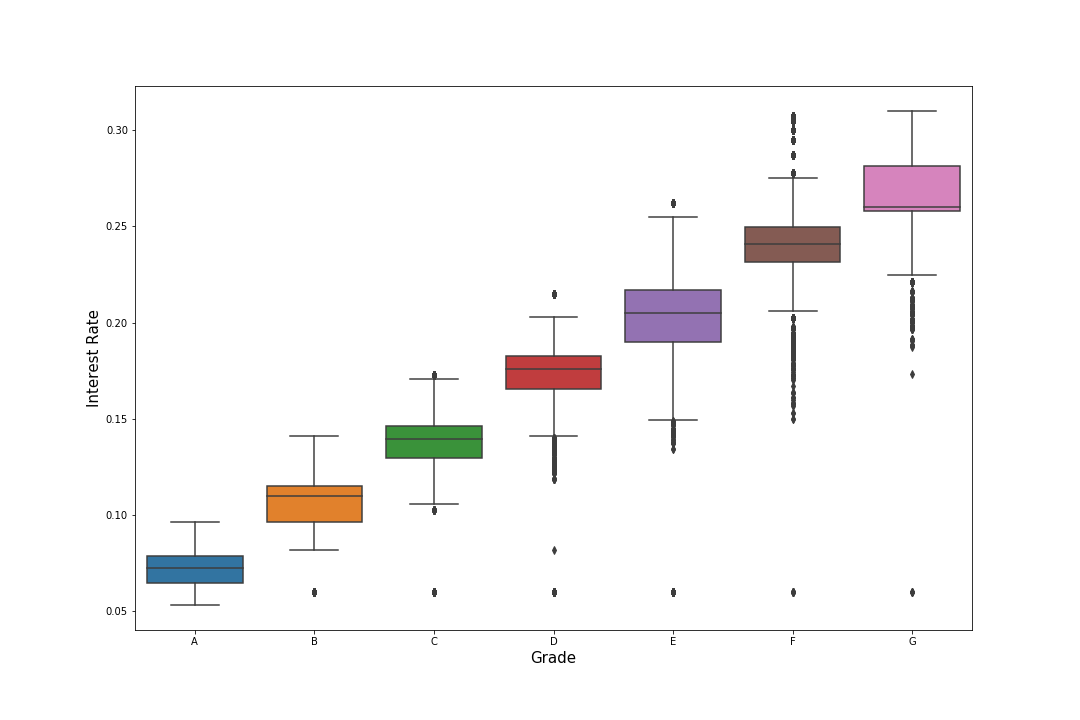}
    \caption{Interest Rates Across Different Grades}
    \label{GradeVSinterest}
\end{figure}

Among the above-mentioned features, some features are very helpful for investors in making decisions. 
For example, the variable ``grade" denotes the grade of each loan that is pre-labeled by Lending Club, ranging from Grade A (the safest) to G (the riskiest). 
The variable ``int\_rate'' denotes the interest rate that is pre-defined by Lending Club.
Figure \ref{GradeVSinterest} shows the interest rate across different grades. 
The Grade A loans are considered to have the lowest PD by Lending Club thus they are associated with the lowest interest rate.
On the other hand, Grade G loans have the relatively higher interest rates due to their high PDs. 
Conservative investors can select relatively ``safer" grades to reduce the investment risk while aggressive investors may select relatively ``riskier" grades to earn a higher profit generated from the higher interest rate. 

Based on these features, researchers of the P2P lending market can use statistical approaches or machine learning methods to help make investment decisions from two perspectives: (1) determining the PDs of the loans and recommending ``safer" loans to investors; and (2) distinguishing profitable loans and recommending ``more profitable" loans to investors \cite{byanjankar2015predicting}.
Again the former approach refers to the credit scoring method while the latter refers to the profit scoring method.

\subsection{Credit Scoring Measure of the Loans} \label{credittarget}
The purpose of credit scoring is to evaluate whether or not the borrowers will repay the loans in a timely manner. In Lending Club, each expired loan (i.e., not during repayment) ends in one of the two states: (1) fully paid; or (2) charged off. 
Fully paid means the borrower has made all the repayments while charged off means the loan is more than 150 days past due. 
We create a target variable named loan\_status using Equation \ref{statusind}.
Table \ref{loan_status_distribute} shows the frequency of each category of loan\_status.
About 80.44\% of the loans are fully paid while 19.56\% of them end with being charged off. 
The credit scoring approach focuses on minimizing the risk of investment by identifying the loans that are fully paid while avoiding those that are charged off. 
Note that for the rest of the paper, we will use the word \textit{default} and \textit{chargedoff} interchangeably.

\begin{equation}\label{statusind}
loan\_status = \begin{cases}
0 &\text{if the loan is fully paid}\\
1 &\text{if the loan is charged off}
\end{cases}
\end{equation}

\begin{table}[htbp]
\renewcommand{\arraystretch}{1.3}
\caption{Distribution of Loan\_status}
\label{loan_status_distribute}
\centering
\begin{tabular}{|p{15mm}|p{15mm}|p{15mm}|p{15mm}|}

            \hline
            Status & Loan\_status & Frequency & Proportion \\
            \hline
            fully paid  & 0 & 904,086 & 80.44\% \\
            charge off  & 1 & 219,809 & 19.56\% \\
                \hline
        \end{tabular}
\end{table}

\begin{figure}[htbp]
\centering
\includegraphics[width=3.3in]{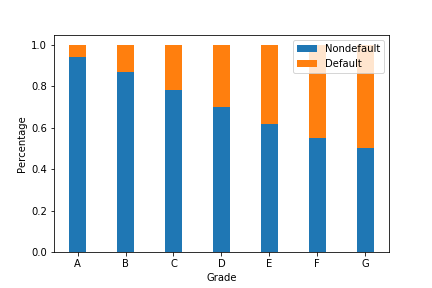}
\caption{Default Rates Across Different Grades}
\label{DefaultRatevsGrade}
\end{figure}

Figure \ref{DefaultRatevsGrade} shows the stacked bar plot of the default rates across different grades labeled by Lending Club. 
As expected, grade A has the lowest default rate while grade G has the highest. 
Therefore, from the perspective of credit scoring, conservative investors should focus on the loans from grade A in order to minimize the default risk. 

\subsection{Profit Scoring Measure of the Loans}\label{profittarget}
The purpose of profit scoring is to evaluate the profit generated by the investment. 
In the Lending Club data, there exists no variable that directly describes the profit of the loans. 
As discussed in Section \ref{introduction}, ARR is an appropriate metric for the profit measure, which can be calculated using the existing features. 
% The ARR formula is described in Equation \ref{ARR_equation}, where Pa is the total payment when the loan expires, Pr denotes the principal, and Y is the number of years of the repayment process. 
For example, suppose a certain investor invests \$6,000 with a nominal interest rate of 14.99\% and 36 scheduled monthly payments. 
Theoretically, if the borrower pays back the loans as scheduled, the ARR is calculated as $(\frac{6000+6000*0.1499*3}{6000})^{\frac{12}{36}} \approx 1.13$.
However, in reality the borrower can pay back earlier or later. 
For example, after 16 months, the borrower may pay back \$7003 including all the principal as well as the interest.
In this case, and the loan expires with the status of being fully paid and the real ARR is calculated as $(\frac{7003}{6000})^{\frac{12}{16}} \approx 1.12$.
Therefore, the investor cannot get the theoretical ARR due to the ``faster" repayment process. 
In our analysis, the real ARR, instead of the theoretical ARR, is used to measure the profit of the loan since the real ARR is the reality that happened in the P2P market. 

\begin{figure}[htbp]
\centering
\includegraphics[width=3.3in]{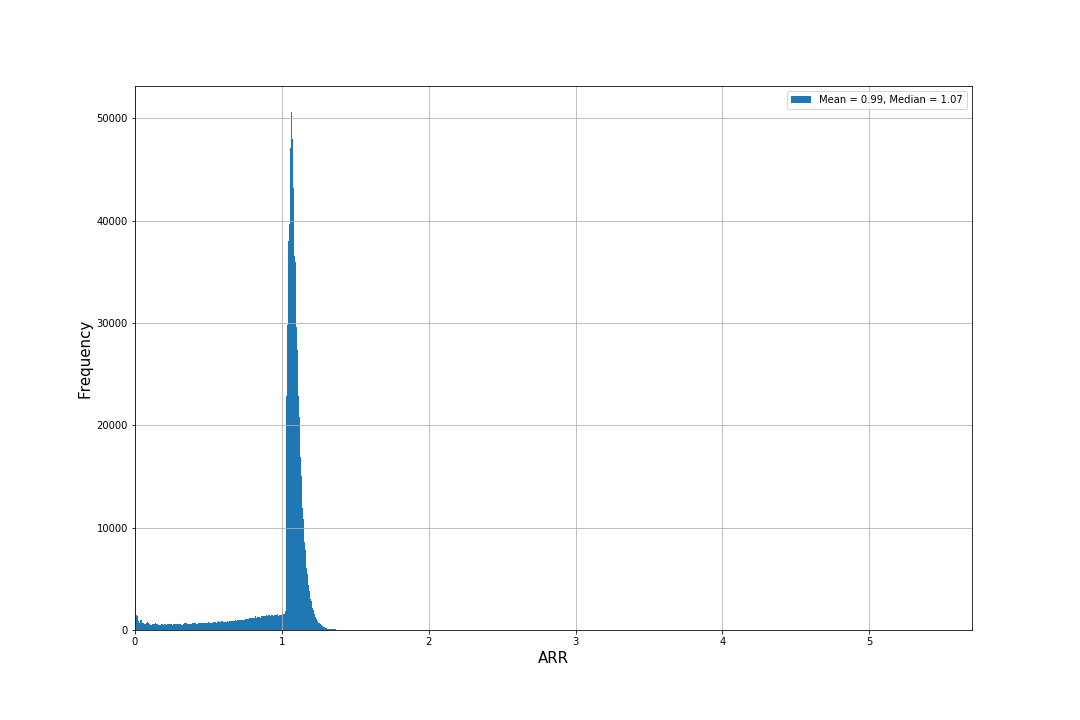}
\caption{Distribution of ARR}
\label{ARR_distribution_hist}
\end{figure}

\begin{figure}[htbp]
\centering
\includegraphics[width=3.3in]{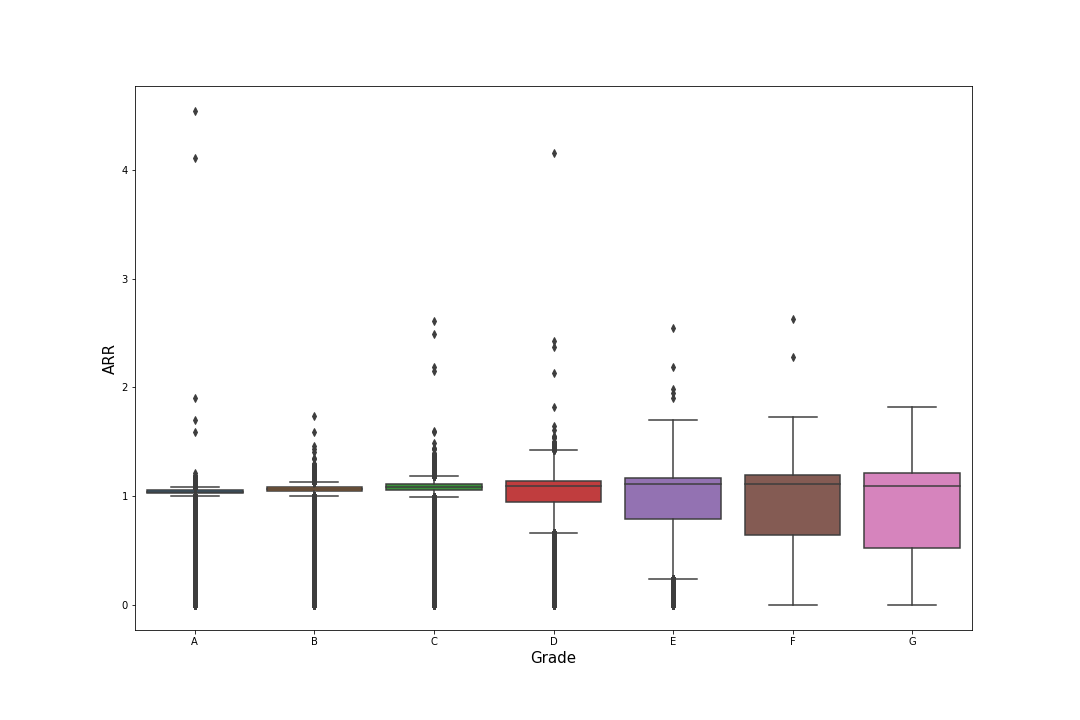}
\caption{ARRs Across Different Grades}
\label{ARRvsGrade}
\end{figure}

Figure \ref{ARR_distribution_hist} shows the distribution of the created variable ARR, which measures the profitability of each loan. 
From Equation \ref{ARR_equation}, we can see that the range of ARR will be $[0, \infty)$.
The minimum value of ARR is 0, which denotes the extremely worst situation where the loan gets zero repayment and the investor loses all the principal.
ARR larger than one denotes a profitable loan, indicating that the borrower pays back more than the principle.
As shown in Figure \ref{ARR_distribution_hist}, the mean and median ARR values are 0.99 and 1.07, respectively.
% Note that in some papers, ARR is defined as $\sqrt[1/Y]{Pa/Pr} - 1$ with the range $(-1, \infty)$.
% In our study, the location of ARR is shifted to the right by one unit on the x-axis in Figure \ref{ARR_distribution_hist} by using the definition in Equation \ref{ARR_equation}.
% The reason to do so is that later in our analysis, we will use ARR to weight different loans during the model training process. 
% Having all non-negative ARR values in the data will avoid the possible confusion during the model training without changing the model performance. \\
Figure \ref{ARRvsGrade} shows the ARRs across different grades. 
The variation of ARR gradually increases from grade A to G. 
Some loans from grades C, D, E, and F lead to a very high ARR, and sometimes even higher than some loans from grades A and B. 
Obviously, the ``safer" loans do not always associate with the ``more profitable" result and the credit scoring approach could not guarantee a good profit.

\subsection{Different Perspectives Between Credit Scoring and Profit Scoring}
\begin{figure}[htbp]
    \centering
    \includegraphics[width=3.3in]{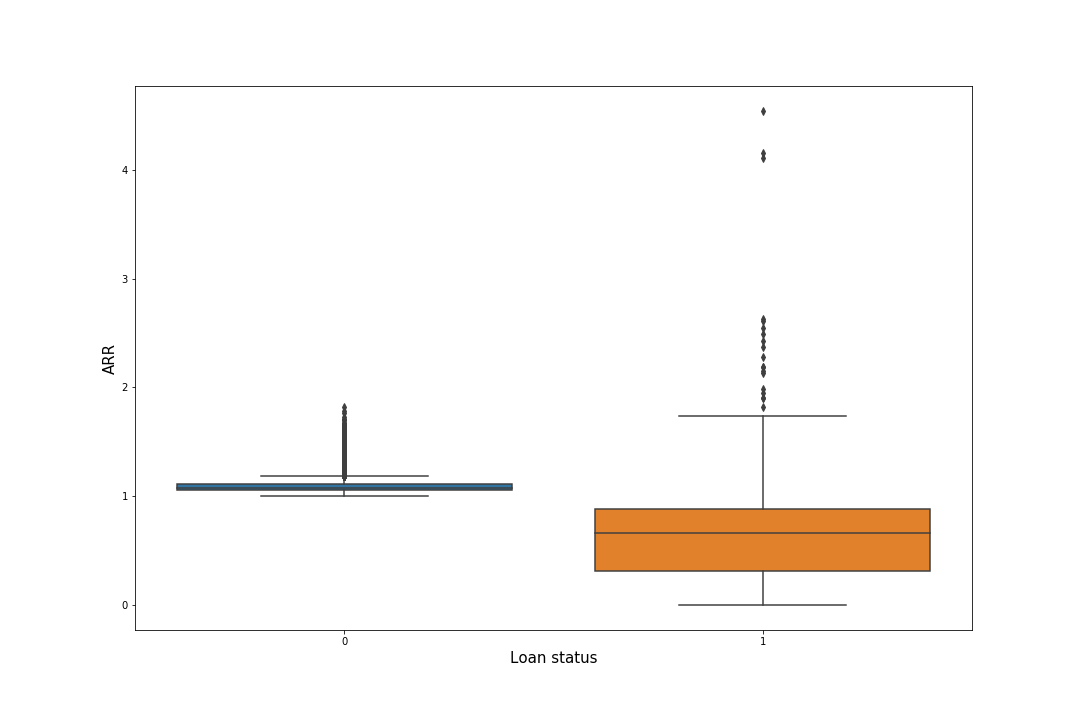}
    \caption{Distribution of ARR Across Different Categories of Loan\_status}
    \label{ARR_distribution}
\end{figure}

As discussed above, credit scoring uses loan\_status as the target variable and aims at predicting the PD of loans. 
Thus, it can evaluate the ``safeness" of loans. 
On the other hand, profit scoring uses ARR as the target variable and aims at predicting the profit of loans. 
Thus, it can evaluate the ``profitability" of loans. 
Figure \ref{ARR_distribution} shows the cross distribution of loan\_status and ARR. 
It is intuitive that there exists a strong relationship between ARR and loan\_status: a defaulted loan (e.g., loan\_status = 1) tends to be associated with a non-profitable ARR and vice versa. 
This can be confirmed by the cross table between ARR and loan\_status shown in Table \ref{crossARRandRisk} and Figure \ref{ARR_distribution}.
As shown by Figure \ref{ARR_distribution}, the variation of ARR of the default loans is much larger than that of the non-default loans, with some default loans resulting in an even higher ARR than non-default loans. 
Consequently, the loans identified with the lowest PD may not always be the best choice for investors, especially for aggressive lenders whose goal is to reach high profitability.
Meanwhile, the default loans with a profitable ARR may be a potential choice for investors but they should be recommended with cautious. 
Moreover, previous studies showed that the explanatory variables differ in predicting loan\_status and profit \cite{serrano2016use}.
Considering all the reasons mentioned above, we conclude that credit scoring and profit scoring measure the loans from different perspectives and one cannot be replaced by another. 
A ``safe" loan identified by the credit scoring approach cannot ensure a ``profit" loan based on a profit scoring approach while a ``profit" loan identified by the profit scoring approach cannot avoid the default risk. 
It is critical to integrate credit scoring and profit scoring together to provide a comprehensive evaluation of the loans, thus may provide better investment decisions.

\begin{table}[htbp]
\renewcommand{\arraystretch}{1.3}
\caption{Cross Table of Loan\_status and ARR}
\label{crossARRandRisk}
\centering
\begin{tabular}{|p{15mm}|p{15mm}|p{15mm}|p{15mm}|}

            \hline
            Loan status & ARR & Frequency & Proportion \\
            \hline
            0 &  $>$ 1 & 904,086 & 80.44\% \\
            1 & $<=$ 1 & 200,859 & 17.87\% \\
            1 & $>$ 1 & 18,950 & 1.69\% \\
                \hline
        \end{tabular}
\end{table}

\subsection{Data Pre-processing}
As discussed above, we finally kept 1,123,895 loans along with 27 variables in the Lending Club data. 
The data is then randomly split into 70\% training set (i.e., 786,726 loans) for the training purpose and a 30\% testing set (i.e., 337,169 loans) for the evaluation purpose. 
For categorical features, a one-hot encoding method is applied.
Take the feature application\_type as an example, which has two categories: Individual and Joint App. 
In our analysis, two columns named application\_type\_Individual and application\_type\_Joint App are created, which are the indicators for the two categories, respectively. 
For numeric features, they are normalized using the min-max normalization to avoid the training bias caused by the various ranges of different attributes \cite{patro2015normalization}.

\subsection{Evaluation Criteria}
The proposed two-stage lightGBM model was first implemented on the Lending Club training set and then evaluated on the test set. 
To confirm that incorporating credit scoring into profit scoring could be beneficial in detecting ``more profitable" loans, we compared its performance with the single profit scoring approach without using any information from credit scoring.
Specifically, two models are compared: an existing profit scoring alone approach based on lightGBM (the One-stage Model), and the proposed two-stage lightGBM method (the Two-stage Model). 
In both models, the hyper-parameters are tuned using the trial and error approach with the goal of minimizing Root Mean Squared Error (RMSE) on the test set. 
The details of the hyper-parameter settings are shown in Table \ref{hyperlightgbm}.
In both models, the final outcome is the predicted value of ARR. 
Loans with a higher predicted ARR would be recommended. 
For the comparison purpose, we compare the profitability of the top 50 loans recommended by the two models in the testing data. 

\begin{table}[htbp]
\renewcommand{\arraystretch}{1.3}
\caption{Hyper-parameter Settings in LightGBM}
\label{hyperlightgbm}
\centering
\begin{tabular}{|p{25mm}|p{30mm}|p{8mm}|}

            \hline
            Name & Description & Value\\
            \hline
            max\_depth & max depth for each tree & 6 \\
            num\_leaves & max leaves for each tree & 10 \\
            feature\_fraction & percentage of features used for each tree & 0.8 \\
            bagging\_fraction & percentage of positive samples used for bagging & 0.5\\
            learning\_rate & shrinkage rate & 0.01\\
                \hline
        \end{tabular}
\end{table}

\section {Experimental Results} \label{resultsanddiscussions}
\begin{figure}[htbp]
\centering
\includegraphics[width=3.3in]{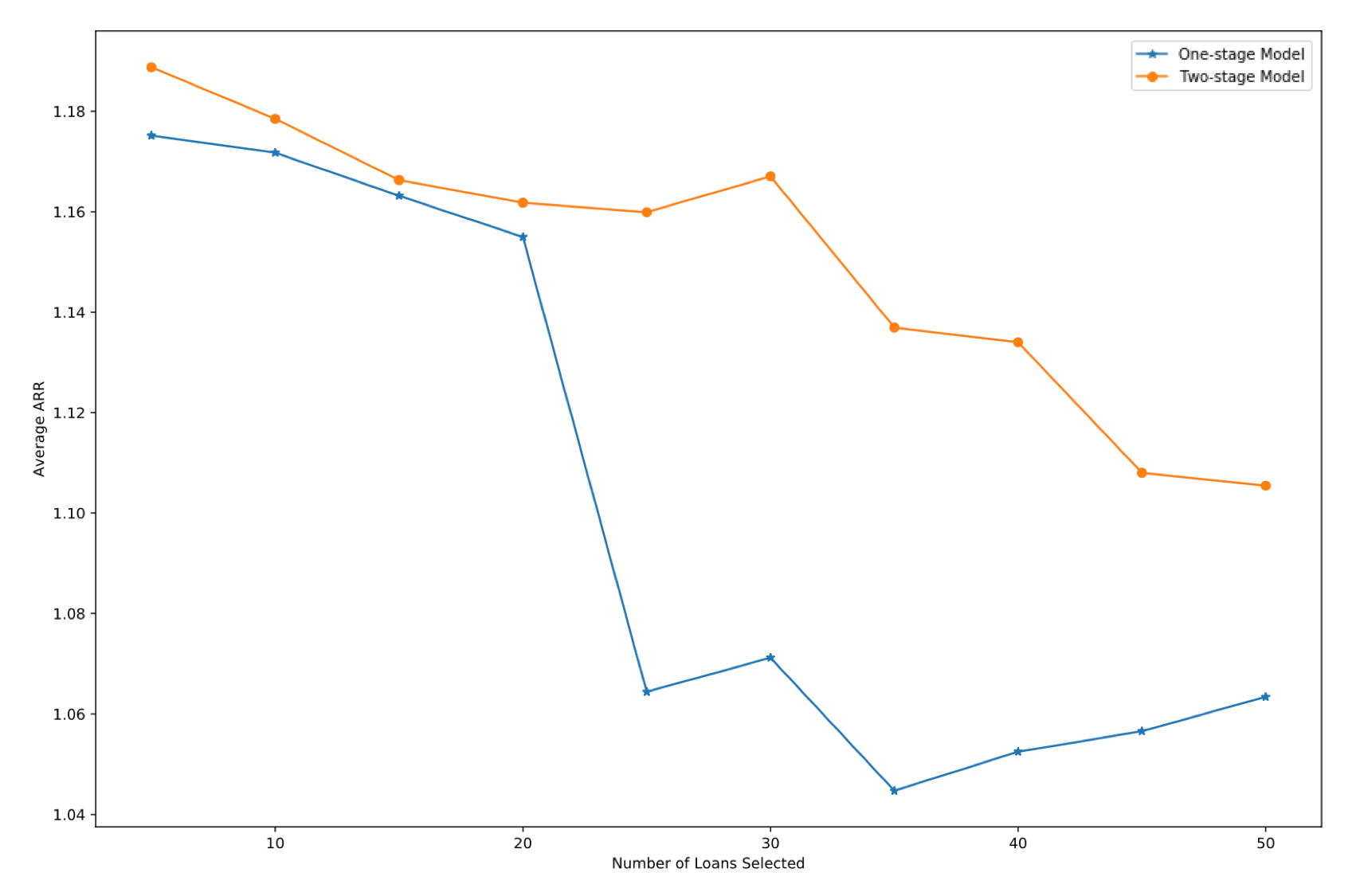}
\caption{Comparison of the One-stage Model and the Two-stage Model in terms of the Average ARR of the Top Loans Selected from the Testing Data}
\label{GBM_compare}
\end{figure}

Figure \ref{GBM_compare} displays the comparison of the average ARR based on the top loans using the two models, where the x-axis denotes the number of top loans identified by the two models changing from 1 to 50. 
Here the value of 50 is big enough for evaluating the model performance since investors tend to care more about the top several (maybe only 5, 10, etc) loans.
It is shown that the profitability of the proposed two-stage model is consistently higher than the profit scoring only method.
The result can strongly confirm our conjecture that incorporating the credit information into profit scoring could be beneficial in identifying the ``more profitable" loans. 
Therefore, the two-stage model would be more preferred in guiding the investment decisions.

To further explore the reason why the two-stage model could detect ``more profitable" loans, 
we compare the constitution of the top 50 loans identified by the two models and the result is shown in Table \ref{top50loans_GBDT}.
Among the top 50 loans identified by the one-stage and the two-stage models, none of them were assigned Grade A by Lending Club.
It can be expected since the safest loans (i.e., assigned by Grade A) can only lead to very small profit because of the low interest rate. 
The one-stage model selects 1 loan from Grade B while the two-stage model didn't select any loan from Grade B. 
Most of the loans recommended by the one-stage model come from Grade D and E while the two-stage model recommends many loans from Grade F. 
Therefore, we can conclude that the two-stage model is more aggressive in selecting loans: it tends to select ``more risky" loans that are defined by Lending Club. 
These risky loans are associated with higher interest rates, thus potentially generate higher profit.
In total, both models have 6 default loans among the top 50 selected loans. 

% \begin{table}[htbp]
% % increase table row spacing, adjust to taste
% \renewcommand{\arraystretch}{1.3}
% \caption{Top 50 Loans Selected by the Two Models}
% \label{top50loans_GBDT}
% \centering
% \begin{tabular}{|p{8mm}|p{13mm}|p{13mm}|p{13mm}|p{13mm}|}
%             \hline
%             Grade & One-stage Model & Default in One-stage Model & Two-stage Model & Default in Two-stage Model \\
%             \hline
%             B & 1 & 0 & 0 & 0 \\
%             C & 7 & 1 & 0 & 0\\
%             D & 14 & 1 & 2 & 0\\
%             E & 14 & 2 & 13 & 1\\
%             F & 10 & 1 & 30 & 3\\
%             G & 4 & 1 & 5 & 2\\
%             \hline
% \end{tabular}
% \end{table}

\begin{table}[htbp]
\renewcommand{\arraystretch}{1.3}
\caption{Top 50 Loans Selected by the Two Models}
\label{top50loans_GBDT}
\centering
\begin{tabular}{|p{8mm}|p{22mm}|p{22mm}|}
            \hline
            Grade & One-stage Model (Default Loans) & Two-stage Model (Default loans) \\
            \hline
            B & 1 (0) & 0 (0) \\
            C & 7 (1) & 0 (0)\\
            D & 14 (1) & 2 (0)\\
            E & 14 (2) & 13 (1)\\
            F & 10 (1) & 30 (3)\\
            G & 4 (1) & 5 (2)\\
            \hline
\end{tabular}
\end{table}

Table \ref{top50loans2_GBDT} summarized the average ARR and the default rate of the top 50 loans identified by the two models.
It shows that the two-stage model can select the loans with much higher ARRs than those selected by the existing one-stage model, which confirms our conjecture that incorporating the credit scoring information is beneficial to improve the performance of the profit scoring approach. 
We have another side result based on Table \ref{top50loans2_GBDT}. 
The default rate generated by the two-stage model is 0.12, which equals to that from the one-stage model.
Therefore, the two-stage model could identify the loans with much higher profits while not introducing extra default risk for investors.

\begin{table}[htbp]
\renewcommand{\arraystretch}{1.3}
\caption{The Average ARR and the Default Rate of the Top 50 Loans Selected by the Two Models}
\label{top50loans2_GBDT}
\centering
\begin{tabular}{|p{18mm}|p{22mm}|p{22mm}|}
            \hline
            Metric & One-stage Model & Two-stage Model\\
            \hline
            Average ARR & 1.09 & 1.13\\
            Default rate & 0.12 & 0.12\\
            \hline
\end{tabular}
\end{table}

\section{Conclusion and Discussion} \label{conclusion}
Profit scoring focuses on profit predictions and it considers the best loans as those with the highest predicted profit. 
The biggest disadvantage of profit scoring is that it ignores the fact that default loans can also be profitable.
In order to overcome the disadvantage of the conventional profit scoring approach, we proposed a two-stage framework that incorporates the credit scoring information into the profit scoring method. 
We used the lightGBM algorithm in both stages 1 and 2 in the model since:
(1) lightGBM is a highly efficient machine learning method in handling large scale data \cite{ke2017lightgbm};
and (2) as one of the state-of-the-art machine learning techniques, lightGBM has not been widely used in the P2P domain, thereby making it necessary to be introduced \cite{ma2018study} \cite{zhang2019research}.
The effectiveness of the proposed two-stage lightGBM is evaluated on the real-world P2P data. 
Results show that compared to a single step profit scoring only method (i.e., the one-stage lightGBM model), the proposed method can identify more profitable loans while it doesn't introduce extra default risk to investors. 
Therefore, it is confirmed that integrating the credit information into profit scoring can provide better investment suggestions to lenders by identifying ``more profitable" loans. 

Different from the previous research which focuses either only on credit scoring or only on profit scoring, this is the first time in our study that a two-stage methodology is proposed with the goal of integrating the two scoring approaches. 
Theoretically, in the future studies, we have many other choices for the classifier in stage 1 and the regressor in stage 2 in the model, as long as the classifiers and regressors could identify the non-linear relationship among the variables. 

The application of the proposed framework is not limited to the P2P area.  It can also be used in other domains that contain two correlated targets. 
Furthermore, the framework can even be extended to a multi-stage workflow to handle problems with multiple targets \cite{yu2008credit}.
Depending on the different data sets and the different research requirements, the best algorithm used in stages 1 and 2 may vary. 
However, the proposed framework can be viewed as the first attempt in the P2P area and demonstrated its promising results.
It may serve as an innovative perspective that could better guide the investment decisions. 

\renewcommand{\shortauthors}{Wang and Ni.}

\bibliographystyle{ACM-Reference-Format}
% %\bibliographystyle{ACM BibTeX style}
\bibliography{sample-base}

% 
% If your work has an appendix, this is the place to put it.
% appendix

\section{Appendices}
The definitions of the variables are shown below: 
\begin{itemize}
   \item Loan characteristics:
   \begin{itemize}
     \item application\_type: A categorical variable denotes whether the loan is an individual application or a joint application with two co-borrowers. 
     \item dti: A.K.A. Debt to Income. A numeric variable denotes the ratio of the borrower's monthly debt to the monthly income. 
     \item grade: A categorical variable denotes the grade of the loan assigned by Lending Club. It ranges from A to G where A is the safest loan and G is the riskiest loan. 
     \item initial\_list\_status: A categorical variable denotes the initial listing status of the loan. 
     \item installment: A numeric variable denotes the monthly payment owed by the borrower. 
     \item loan\_amnt: A numeric variable denotes the total amount of money of a loan. 
     \item purpose: A categorical variable denotes the purpose for the loan. 
     \item sub\_grade: A categorical variable denotes the subgrade of the loan assigned by Lending Club. It ranges from A1 to G5 where A1 is the safest loan and G5 is the riskiest loan.
     \item term: A categorical variable denotes the term of the loans. It can be either 36 months or 60 months. 
     \item verification\_status: A categorical variable denotes whether the income of the borrower was verified or not. 
   \end{itemize}
   
   \item Credit worthiness:
       \begin{itemize}
        \item acc\_now\_delinq: A numeric variable denotes the number of accounts on which the borrower is now delinquent. 
        \item deling\_2yrs: A numeric variable denotes the number of delinquencies the borrower had in the past two years. 
        \item cr\_line\_month: A numeric variable denotes the credit age of the borrower (in months) from the earliest credit trade line listed in the credit report to the date when the loan is applied. 
        \item fico\_range\_high: A numeric variable denotes the upper boundary of the borrowers FICO score range when the loan was originated. 
        \item fico\_range\_low: A numeric variable denotes the lower boundary of the borrowers FICO score range when the loan was originated. 
        
        \balance 
        
        \item inq\_last\_6mths: A numeric variable denotes the number of inquiries listed in borrower's credit report during the past 6 months. 
        \item open\_acc: A numeric variable denotes the number of open trade lines in the borrower's credit report. 
        \item pub\_rec: A numeric variable denotes the number of derogatory in the borrower's credit report. 
        \item revol\_bal: A numeric variable denotes the total credit revolving balance.
        \item revol\_util: A numeric variable denotes the amount of revolving credit limit that the borrower currently has. 
        \item total\_acc: A numeric variable denotes the total number of open credit accounts on the borrower's credit file. 
     \end{itemize}
     
   \item Borrower information:
       \begin{itemize}
        \item addr\_state: A categorical variable denotes the state of the address provided by the borrower in the loan application.
        \item annual\_inc: A numeric variable denotes the annual income information provided by the borrower. 
        \item emp\_length: A numeric variable denotes the length of time in years the borrower is employed in a company.
        \item emp\_title: A categorical variable denotes the job title provided by the borrower when applying for the loan.
        \item home\_ownership: A categorical variable denotes whether the borrower owns the house.
        \item zip\_code: A categorical variable denotes the first three digits of the zip code provided by the borrower in the loan application.
     \end{itemize}
\end{itemize}

\end{document}